\documentclass[aps,amsmath,amssymb,superscriptaddress,reprint,floatfix,longbibliography]{revtex4-1}
\usepackage{graphicx}% Include figure files

\usepackage{bm}
\usepackage{amsmath,amssymb}

\usepackage[hidelinks]{hyperref}
\usepackage{siunitx}
\usepackage{physics}

\begin{document}

\title{Quantum correlation measurement with single photon avalanche diode arrays}
\author{Gur Lubin}\thanks{These authors contributed equally to this work}\affiliation{Department of Physics of Complex Systems, Weizmann Institute of Science, Rehovot, Israel}
\author{Ron Tenne}\thanks{These authors contributed equally to this work}\affiliation{Department of Physics of Complex Systems, Weizmann Institute of Science, Rehovot, Israel}
\author{Ivan Michel Antolovic}\thanks{These authors contributed equally to this work}\affiliation{School of Engineering, École Polytechnique Fédérale de Lausanne (EPFL), Neuchâtel, Switzerland}
\author{Edoardo Charbon}\affiliation{School of Engineering, École Polytechnique Fédérale de Lausanne (EPFL), Neuchâtel, Switzerland}
\author{Claudio Bruschini}\email{claudio.bruschini@epfl.ch}\affiliation{School of Engineering, École Polytechnique Fédérale de Lausanne (EPFL), Neuchâtel, Switzerland}
\author{Dan Oron}\email{dan.oron@weizmann.ac.il}\affiliation{Department of Physics of Complex Systems, Weizmann Institute of Science, Rehovot, Israel}

\begin{abstract}
Temporal photon correlation measurement, instrumental to probing the quantum properties of light, typically requires multiple single photon detectors. Progress in single photon avalanche diode (SPAD) array technology highlights their potential as high performance detector arrays for quantum imaging and photon number resolving (PNR) experiments. Here, we demonstrate this potential by incorporating a novel on-chip SPAD array with 55\% peak photon detection probability, low dark count rate and crosstalk probability of 0.14\% per detection, in a confocal microscope. This enables reliable measurements of second and third order photon correlations from a single quantum dot emitter. Our analysis overcomes the inter-detector optical crosstalk background even though it is over an order of magnitude larger than our faint signal. To showcase the vast application space of such an approach, we implement a recently introduced super-resolution imaging method, quantum image scanning microscopy (Q-ISM).
\end{abstract}

\maketitle

%%%%%%%%%%%%%%%%%%%%%%%%%%  body  %%%%%%%%%%%%%%%%%%%%%%%%%%
\section{Introduction}
\begin{figure}[t]
\centering
\includegraphics[width=.9\linewidth]{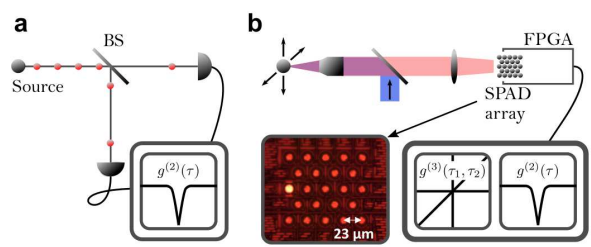}
\caption{\label{fig:intro}\textbf{Photon correlation setups.} \textbf{a.}~Hanbury Brown and Twiss intensity interferometer. A beamsplitter splits the incident photons to two correlated detectors. A single photon emitter source will be characterized by a dip at the zero time delay of the second order photon correlation. Higher order correlations demand more beamsplitters and detectors. \textbf{b.}~The SPAD array photon correlation setup. A SPAD array is positioned at the image plane of a scanning confocal microscope, resulting in the splitting of the beam by diffraction onto 23 detectors. \textbf{Inset.}~An optical image of the SPAD array.}
\end{figure}

Quantum imaging is an emerging field in optical microscopy attempting to overcome the classical limitations of imaging in terms of precision and spatial resolution\cite{Brida2010a,Ono2013a,Schwartz2013,Israel2014a,Tsang2015,Unternahrer2018}. While quantum imaging methods differ in the illumination, imaging optics and data analysis procedures, they all rely on characterization of a quantum state of light at the imaging port\cite{Ono2013a,Israel2017,Classen2017,Toninelli2019}. A critical component for a quantum microscope, therefore, is an imaging detector capable of analyzing quantum signatures of the output light. Measuring the second order correlation with a Hanbury-Brown and Twiss (HBT) intensity interferometer is the prevailing method to probe non-classical properties of light\cite{Gerry2005}. In a standard setup a beam is equally split onto two detectors and their outputs are correlated (\autoref{fig:intro}(a)). However, the standard HBT setup is inherently a single pixel measurement not suitable for widefield quantum imaging methods.

An alternative approach, compatible with imaging techniques, uses the diffraction limit of the imaging system to split the optical signal between several detectors in an array, and correlates their output\cite{Schwartz2013,Israel2017,Classen2017,Tenne2018} (\autoref{fig:intro}(b)). While commercial low light cameras such as intensified cameras and electron multiplying charge coupled devices (EMCCD) are natural candidates to perform such tasks, they operate only at relatively low $\sim\si{\kilo\hertz}$ frame rates. Measuring photon correlation is a single-shot-per-frame experiment, i.e. your signal level is at most a single reading (e.g. a simultaneous photon pair) per frame per diffraction limited spot\cite{Schwartz2013,Unternahrer2016,Toninelli2019}. As a result, an imaging detector with a $\sim\si{\mega\hertz}$ readout rate is extremely beneficial to acquire the quantum contrast within reasonable exposure times.

Over the past two decades, progress in complementary metal–oxide–semiconductor (CMOS) processing of single photon avalanche diode (SPAD) array technology has positioned them at the forefront of time resolved imaging\cite{Lussana2015,Antolovic2017,Castello2019}. Advancements in array dimensions, detection efficiency and low dark count rates (DCR), along with the inherent single photon sensitivity and sub nanosecond time resolution, enable a plethora of low light level and time-resolved applications\cite{Bruschini2019a}. A CMOS SPAD array is therefore a natural candidate for imaging quantum correlations\cite{Agnew2016}, especially for low light applications such as biological microscopy.

In addition to the single pixel limitation, a standard HBT setup is also restricted to the measurement of two simultaneous photons at the most. Measurement of higher photon numbers requires extending the HBT scheme to include multiple detection ports and becomes quite cumbersome\cite{Eisaman2011}. Therefore, characterizing the photon number distribution, termed photon number resolving (PNR) detection, is a challenge that can also benefit from performing HBT measurements with a CMOS SPAD array. PNR detection schemes can be classified into two categories: a single PNR detector and multiplexed (temporally or spatially) single photon detectors. Single detector PNR schemes, such as visible light photon counters (VLPC)\cite{Kim1999,Waks2003}, superconducting transition edge sensors (TES)\cite{Cabrera1998,Miller2003} and nano-structured transistor devices\cite{Luo2018,Gansen2007}, rely on the proportionality of the output signal to the number of photons. While these techniques allow high efficiency detection with very low noise levels\cite{Lita2008}, they demand cryogenic cooling\cite{Eisaman2011}, have a limited saturation rate\cite{Hadfield2009} and often require optical coupling through a cavity which limits their usefulness for spatially and spectrally multimode signals\cite{Lita2008}. Time multiplexing of a single photon detector is typically achieved by splitting the signal to different fiber delay lines\cite{Achilles2003,Fitch2003,Micuda2008}. Although this approach enables PNR with only one inexpensive single-photon detector, it requires the use of very long fibers and is currently limited to single spatial mode signals. Finally, spatial multiplexing can be achieved by utilizing a two dimensional detector array, such as a CMOS SPAD array, and the diffraction of light as a natural beamsplitter onto an arbitrarily large number of detectors\cite{Divochiy2008,Eraerds2007,Jiang2007,Unternahrer2016}.

However, implementation of both PNR detection and quantum imaging techniques with a CMOS SPAD array requires overcoming the effect of the characteristic crosstalk between neighbouring detectors in the array\cite{Rech2008}. While crosstalk has a negligible effect on intensity measurements, it directly competes with the short-time photon correlation signal, and is typically of a much larger scale.

In this work, we use a novel 23 pixel SPAD array (see \autoref{fig:intro}(b) inset and ref \cite{Antolovic2018}) with minimized crosstalk, fabricated in CMOS image sensor technology, to measure photon correlations from faint sources by statistically compensating for crosstalk artifacts. To demonstrate the PNR capabilities of the detector array we measure second and third order photon antibunching in the photoluminescence of a single quantum dot. By placing the detector array in the imaging plane of a confocal microscope we were able to implement quantum image scanning microscopy (Q-ISM), a recently introduced quantum imaging technique that was already demonstrated in biological imaging\cite{Tenne2018}, with a substantially simplified detection scheme.

\section{Crosstalk characterization}\label{sec:ct}
\begin{figure*}[t]
\centering
\includegraphics[width=.9\textwidth]{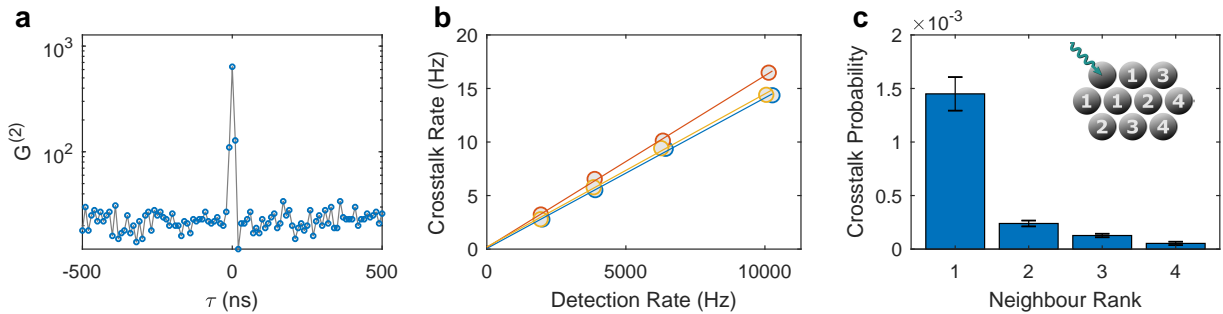}
\caption{\label{fig:ct}\textbf{SPAD array crosstalk characterization.} \textbf{a.}~Typical second order correlation of photon arrival times for two nearest-neighbour detectors in the array, in response to homogeneous illumination by a thermal source (analyzed from $10^7$ detections over \SI{\sim43}{\second}). Note the sharp peak at zero time delay attributed to crosstalk. \textbf{b.}~Crosstalk linearity in detection rate. Each colored set of markers represent a different nearest neighbour detector pair. Lines of corresponding color represent a linear fit for each pair. \textbf{c.}~Characterization of crosstalk dependence on detector distance. Each bar shows the crosstalk probability averaged over all detector pairs at a certain distance. Error bars represent one standard deviation of the distribution over these pairs. The values suggest that crosstalk is significant mostly for nearest neighbour detectors. \textbf{Inset.}~Visualization of neighbour rank. A photon (green arrow) is absorbed in the upper left detector. The neighbouring detectors are ranked by distance from the excited detector. Nearest neighbours are rank 1.}
\end{figure*}

Recent progress in the design and manufacturing processes of CMOS based SPAD arrays led to sub-megapixel arrays\cite{Ulku2019}, low DCR and improved photon detection efficiency (PDE)\cite{Veerappan2014,Gyongy2018}. This has been achieved by a synergy between innovative SPAD designs, process improvements and 3D integrated circuit (IC) technology advancements. In this work we focus on small arrays, optimized for confocal microscopy consisting of \num{23} pixels positioned in a 2D hexagonal lattice with a period of \SI{23}{\micro\meter}, feeding a field-programmable gate array (FPGA)\cite{Antolovic2018}. These SPAD arrays feature an average room-temperature DCR lower than 100 counts per second (cps) per pixel. On average, less than 2\% of the pixels are considered hot - with a DCR of over \SI{1000}{cps} (the array used in this work had one hot pixel). The maximum measured count rate with passive recharge is limited to around \SI{10}{Mcps} per pixel, the afterpulsing probability is 0.1\% and the dead time is \SI{\sim100}{\nano\second} (at a passive quenching resistance of 50-~\SI{500}{\kilo\ohm}). For the detector used here we estimated the photon detection efficiency (PDE), by comparison with an independent EMCCD measurement, to be 42\% at \SI{515}{\nano\meter} and \SI{7}{\volt} excess bias. While PDE, DCR, dynamic range and afterpulsing are often depicted in the literature, only few works discuss the issue of inter-pixel crosstalk\cite{Rech2008,Aull2015,Ficorella2016,Jahromi2018,Wu2018}. While both afterpulsing and crosstalk generate artificial correlations, afterpulsing artifacts are avoided here altogether by disregarding the autocorrelation of any single detector, as typically done in HBT experiments. 

SPAD crosstalk can be both electrical and optical. Electrical crosstalk can be caused by charge diffusion from electronics and adjacent pixels. This effect is eliminated by using substrate isolated SPAD designs\cite{Veerappan2014}. Optical crosstalk is caused by spontaneous photon emission within the few nanoseconds avalanche duration, detected by another detector in the array\cite{Rech2008}. It can be minimized by reducing the amount of charge flow through the SPAD by active quenching and implementing opaque deep trench isolation around the SPAD. In such an optimized SPAD implementation, optical crosstalk occurs only by photon scattering from structures around the SPAD, usually metal connections. In the remainder of this section we describe in detail the characterization procedure of the inter-pixel optical crosstalk and its results.

To characterize crosstalk, the SPAD array was illuminated with spatially homogeneous, white light illumination, produced by a halogen lamp. The thermal state of light generated by the lamp leads to positive correlations (photon bunching) at the scale of the coherence time, $\tau_c\approx\SI{10}{\femto\second}$, much shorter than the FPGA timing resolution ($t_{clk}=\SI{10}{\nano\second}$) and the SPAD jitter (\SI{\sim 120}{\pico\second}). Correlations measured with a $t_{clk}$ temporal resolution should thus present only a minute deviation from those of a classical coherent state, $10^{-6}$ with respect to the signal, well below the noise. As a result, we can treat this light source as effectively uncorrelated, and the second order correlation ($G^{(2)}(\tau)$) of photon arrival times for such a source should result in a flat line. However, as is evident from \autoref{fig:ct}(a), the $G^{(2)}$ of two neighbouring detectors in the array shows a distinct peak at a zero time delay (for a detailed description of the $G^{(2)}$ data analysis see \autoref{app:analysis}). These extra photon pairs are attributed to inter-detector optical crosstalk. Despite the short time scale of optical crosstalk, due to small differences in clock timings for different detectors, some positive correlation is also present at $\tau=\pm 10\ \si{\nano s}$ (see \autoref{fig:ct}(a)). Since the magnitude of this effect is much smaller than the excess correlation at zero delay, we neglect this contribution in the following analysis.

The above mentioned mechanism for optical crosstalk implies linearity of the crosstalk correlation term with respect to the number of detections, up to the detector saturation effects.  To test this linearity of optical crosstalk, \autoref{fig:ct}(b) presents the crosstalk rates for three different detector pairs \textit{versus} the rate of detected photons (circles). Linear fits of the data (lines of corresponding color) show that at illumination levels well below the detector saturation, optical crosstalk is indeed linear with the number of detections. We can therefore define the crosstalk probability $p^{CT}_{i,j}$ as the probability that a detection in pixel $i$ will lead to a false detection at detector $j$. The crosstalk probabilities for each detector pair can be inferred from the $G^{(2)}$ analysis of a single `classical light' measurement according to
\begin{equation}
    p_{i,j}^{CT}=\frac{ G^{(2)}_{i,j}(0)-\left\langle G^{(2)}_{i,j}(\infty) \right\rangle }{ n_i + n_j }, 
\end{equation}
where $\langle G^{(2)}_{i,j}(\infty) \rangle$ is an average of $G^{(2)}_{i,j}$ (the second order correlation of pixels $i$ and $j$) excluding \numlist{-1;0;1} clock delays and $n_k$ is the total number of photons measured in detector $k$. Note that we assume here that crosstalk probabilities are symmetric to the exchange of $i$ and $j$; i.e.\ the probability of a photon detection in pixel $i$ resulting in a crosstalk detection in pixel $j$ is equal to that of a photon detection in pixel $j$ leading to a crosstalk detection in pixel $i$. 

\autoref{fig:ct}(c) presents the mean values of these probabilities over all pairs at four inter-detector distances in the array. The probability for a false detection pair is $\num[separate-uncertainty = true, tight-spacing = true]{1.45 \pm 0.16 e-3}$ for nearest neighbor pixel pairs, while the corresponding value for next nearest neighbors is lower by a factor of $\sim6$. Note that these values are well below the typical values reported in the literature for SPAD arrays ($\sim 1\%$)\cite{Aull2015,Ficorella2016,Jahromi2018,Wu2018}, while conventional SiPM detectors usually have higher probabilities of $4-20\%$\cite{Otte2017}. Compared to SiPMs, SPAD arrays usually have a smaller capacitance and consequently a smaller charge flowing through the SPAD. In our design, the SPAD is capacitively isolated from the rest of the circuit by a pixel-level inverter. Additional experiments, presented in \autoref{app:stability}, show that there are no noticeable short or long-term temporal variations in the crosstalk probabilities.

\section{Second order photon antibunching}
\begin{figure*}[t]
\centering
\includegraphics[width=.9\textwidth]{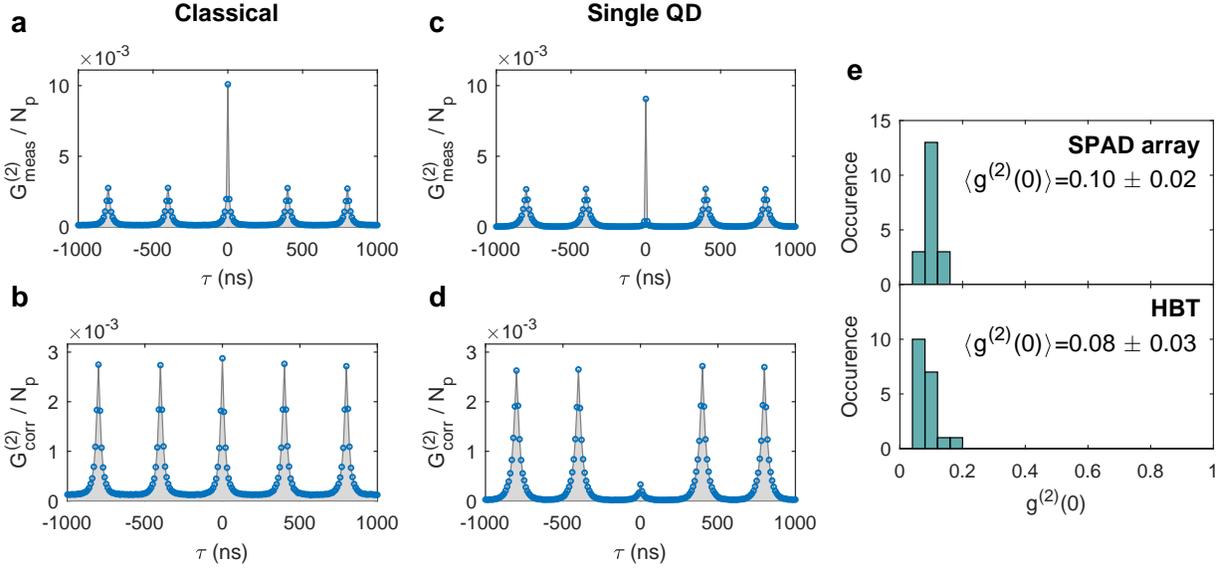}
\caption{\label{fig:g2}\textbf{Second order photon correlations.} Second order photon arrival time correlations ($G^{(2)}(\tau)$) for: \textbf{a.}~Classical light (large ensemble of QDs), \textbf{b.}~same as \textbf{a} after crosstalk correction, \textbf{c.}~single QD and \textbf{d.}~single QD after crosstalk correction. All values are normalized by the total number of measured photons ($N_p$). The correlation peaks are centered at integer multiples of the laser inter-pulse separation (400\si{\nano s}) broadened by the QDs emission lifetime (\SI{\sim26}{\nano\second}). In panel a and c, the high crosstalk peaks at zero time delay can be clearly seen. In panel b the zero delay peak is featureless with respect to non-zero peaks, as expected for classical light sources following Poissonian statistics. The zero delay peak in panel d is significantly lower than the non-zero delay peaks, as expected from an antibunched light source. Classical and single QD correlation curves were analyzed from $10^7$ detections over \SI{\sim103}{\second} and \num[separate-uncertainty = true, tight-spacing = true]{8e6} detections over \SI{\sim105}{\second} respectively. \textbf{e.}~Histograms of the normalized second order correlation function at zero time delay ($g^{(2)}(0)$) after crosstalk correction for 19 single QD measurements, utilizing the SPAD array (top panel) and on a similar sample with a standard HBT setup using two separate SPADs and a split optical fiber (bottom panel). Note the very good agreement of the distributions' mean value and width. The mean value is much smaller than $0.5$, indicative of single photon emitters.}
\end{figure*}
While optical crosstalk generates unwanted correlations at short time scales, its linear dependence on the number of detections allows us to pre-characterize it and subtract an estimated correction from any photon correlation measurement. In the following, we employ the SPAD array as a confocal microscope detector in order to test its ability to characterize a quantum state of light in a photon-starved microscopy application by applying a crosstalk correction to the measured $G^{(2)}$ function. 

For this purpose we have built a custom confocal setup around a commercial inverted microscope (Eclipse T\textit{i}-U, Nikon). A pulsed diode laser (LDH-P-C-470B, PicoQuant) provides a collimated beam at a wavelength of \SI{470}{\nano\meter} and a repetition rate of \SI{2.5}{\mega\hertz}. The beam is focused by a high numerical aperture (NA) oil immersion objective lens ($\times$100, 1.3~NA, Nikon) which also collects the resulting fluorescence light. Back-scattered laser light is filtered out by a dichroic mirror (505~LP, Chroma) and a long pass dielectric filter (488~LP, Semrock). Finally a relay lens images the fluorescence onto the SPAD array with a total magnification of $\times190$, so that the full width at half maximum (FWHM) of the point spread function (PSF) corresponds to $\sim2.8$ pitch periods in the detector plane, and over 96.5\% of the collected light falls within the detector array.

As an initial test, we generate classical light, following Poissonian statistics, by exciting a drop of a dense solution of core/shell/shell CdSe/CdS/ZnS quantum dots (QDs) (see \autoref{app:samp_prep}). In order to estimate the second order correlation function, $G^{(2)}(\tau)$, we histogram photon pairs from the entire detector array according to the time difference between the two detections. An analysis of a photon trace from the whole SPAD array produced during a \SI{\sim103}{\second} exposure is shown in \autoref{fig:g2}(a). At non-zero time delays one can observe correlation peaks centered at integer multiples of the laser inter-pulse separation broadened by the QD’s emission lifetime. In contrast, a narrow crosstalk peak is dominant at zero time delay. Although the probability for crosstalk is much smaller than unity, this peak overwhelms the photon correlation features since the occurrence of crosstalk is more probable than that of two separate photon detections in any specific time delay. To correct for the effect of optical crosstalk and estimate the light-only second order correlation we define the corrected second order correlation function $G^{(2)}_{corr}$ as follows:
\begin{equation}
    G^{(2)}_{corr}(\tau) \equiv 
    \begin{cases}
        G^{(2)}_{meas}(0) - \displaystyle\sum_{i\neq j}n_i\cdot p^{CT}_{i,j} &  \quad\tau=0 \\
        G^{(2)}_{meas}(\tau) & \quad\tau\neq0
    \end{cases},
\end{equation}
where $G^{(2)}_{meas}$ is the as-measured correlation function, $G^{(2)}_{corr}$ is the corrected correlation function excluding crosstalk effects and the summation is over all detector pairs excluding the diagonal terms ($i=j$). The correction term for the zero delay point applies the pre-characterized crosstalk probabilities $p^{CT}_{i,j}$ discussed in \autoref{sec:ct}.

The corrected second order correlation function ($G^{(2)}_{corr}$), shown in \autoref{fig:g2}(b), presents a featureless peak at zero time delay similar in height and width to the neighboring peaks. The normalized value of the second order correlation function, $g^{(2)}(0)$, is estimated as the ratio between the area under the zero time delay peak and the mean area under all other peaks in $G^{(2)}_{corr}$ (see \autoref{app:analysis}). It matches the expected value of 1 for classical light, deviating by less than 0.1\%. The agreement with theory indicates that with appropriate crosstalk correction the SPAD array performs well as an HBT setup in low light conditions.

To demonstrate the applicability of an on-chip SPAD array as a detector of quantum light, we measure individual QDs sparsely dispersed in a spin coated film on a glass cover slip (see \autoref{app:samp_prep}). QDs are well-known as single-photon-at-a-time emitters; the emission of two photons within the same radiative lifetime is strongly inhibited\cite{Michler2000}. \autoref{fig:g2}(c) presents a photon correlation analysis of such a measurement. As in \autoref{fig:g2}(a), here too, the zero delay crosstalk feature is the most prominent one. However, once the crosstalk estimate is subtracted (\autoref{fig:g2}(d)), we can notice that the correlation peak around zero delay is considerably lower than the non-zero delay peaks, as expected from an antibunched source of light. \autoref{fig:g2}(e) shows the distribution of $g^{(2)}(0)$ for 19 single QD measurements in two measurement setups. The histogram in the top panel was measured with the SPAD array setup, while the bottom in a standard HBT experiment, employing two commercial SPAD detectors (COUNT-20B, Laser Components) and a split optical fiber as a beamsplitter. The average error in the estimate of $g^{(2)}(0)$ for individual QDs is $\approx0.008$ and $\approx0.004$ for the SPAD array and standard HBT measurements, respectively; much smaller than the distributions' standard deviations. We therefore interpret the antibunching distributions as a sample of the various values within the synthesis products. The agreement between the mean and standard deviation values measured with the two setups suggests that properly corrected SPAD array data does not introduce an appreciable bias to the $g^{(2)}(0)$ measurement. The mean value is much smaller than $0.5$, indicative of single photon emitters.

\begin{figure*}[t]
\centering
\includegraphics[width=.7\textwidth]{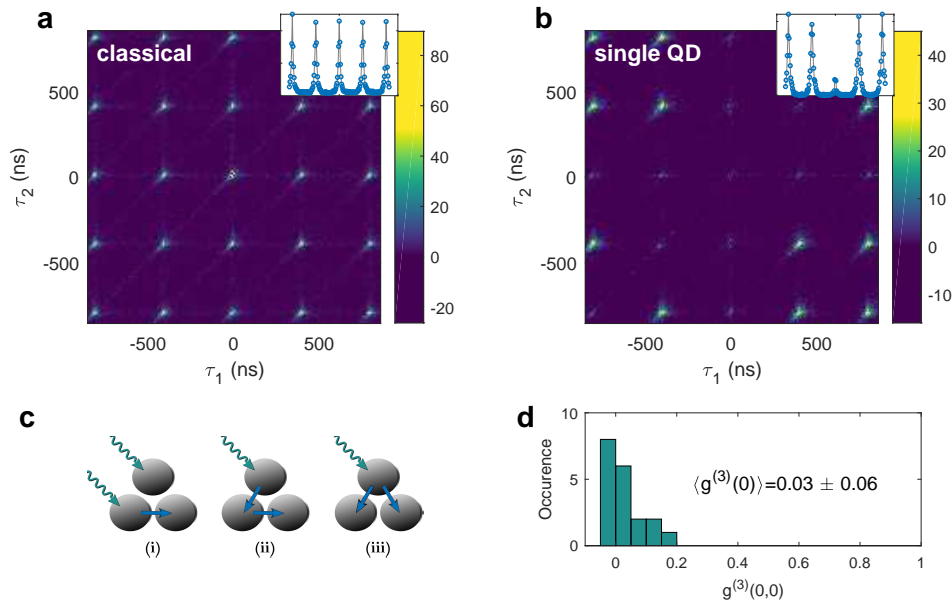}
\caption{\label{fig:g3}\textbf{Third order photon correlations.} Crosstalk corrected third order photon correlations for: \textbf{a.}~Classical light (large ensemble of QDs), \textbf{b.}~single QD. Photon triplets are histogramed according to the difference in their arrival times, $\tau_1$ and $\tau_2$ being the delays between the arrival of one (randomly selected) photon of the triplet and the arrival times of the other two photons respectively. The colorbar represents number of triplets in histogram bin ($t_{clk}=\SI{10}{\nano s}$ binning in both axes). Negative values are a result of crosstalk over-correction due to noise. The observed grid of peaks corresponds to the \SI{2.5}{\mega\hertz} frequency of the pulsed excitation. The peaks' profile match the fluorescence lifetime of the QDs. Note the decimation of peaks along the two axis and one of the diagonals in panel b, indicating photon antibunching (low $g^{(2)}(0)$). Insets are the second order correlation estimations attained by full vertical binning of the $G^{(3)}$ values. \textbf{c.}~The three possible pathways for crosstalk to form false $G^{(3)}(0,0)$ triplets from incident photons (green arrows) and crosstalk events (blue arrows). \textbf{d.}~Histogram of crosstalk corrected $g^{(3)}(0,0)$ values from 19 different QDs.}
\end{figure*}

\section{Third order photon antibunching}
Realizing an HBT setup with an on-chip SPAD array offers flexibility and scalability in the measurement of photon correlations. An example of this advantage is the possibility of measuring photon correlation of orders higher than two. A measurement of the third order photon correlation would typically require upgrading the experimental setup to include further optical elements, detectors and time-to-digital converting channels\cite{Ta2010,Stevens2014}. An on-chip SPAD array used in a confocal setup offers the opportunity to split the light between multiple channels without any modifications to the experimental setup. In fact, the same data set used to produce the $G^{(2)}_{corr}(\tau)$ curves shown in \autoref{fig:g2}(b) and \ref{fig:g2}(d) is used to analyze the third order correlation function, $G^{(3)}_{corr}(\tau_1,\tau_2)$, shown in \autoref{fig:g3}(a) and \ref{fig:g3}(b) respectively. To generate these figures, we histogram photon triplets according to the difference in their arrival times, $\tau_1$ and $\tau_2$ being the delays between the arrival of one (randomly selected) photon of the triplet and the arrival times of the other two photons respectively.

Note that the triplets lying on the $\tau_1=0$, $\tau_2=0$ and $\tau_1=\tau_2$ include two simultaneous detections whereas the origin point $\tau_1=\tau_2=0$ contains three detection within $t_{clk}$. In order to correctly evaluate $G^{(3)}_{corr}(\tau_1,\tau_2)$ at the above mentioned time bins, it is imperative to subtract the contribution of crosstalk at these points. For brevity, we leave the full mathematical form of this correction for \autoref{app:ct_cor}, and qualitatively describe in the following the different crosstalk terms that need to be accounted for. At the origin point, in particular, one has to consider three types of crosstalk events leading to false positive detection triplets, schematically shown in \autoref{fig:g3}(c). In the first type, following a coincidental pair of photo-induced avalanches, crosstalk from one of them may lead to a detection in a third detector (i). Additionally, an event in which only one of the three detections is due to a signal photon can occur in one of two ways, termed here serial (ii) and  parallel (iii). In a serial event, a detected photon leads to crosstalk detection in a second detector, which in turn results in crosstalk detection at a third detector. A parallel third order crosstalk event consists of a single photo-detection leading to the emission of light detected by two neighboring pixels.

After proper subtraction of the estimate of all these contributions, we obtain the crosstalk corrected $G^{(3)}_{corr}$ function (\autoref{fig:g3}). Since a single QD preferentially emits only one photon at a time, peaks centered on the axes and the $\tau_1=\tau_2$ diagonal are highly attenuated (by a factor of $g^{(2)}(0)$), as seen in \autoref{fig:g3}(b). The further attenuation of the zero delay peak around the origin suggests that the detection of three simultaneous photons is even less probable than that of a simultaneous photon pair. While third-order antibunching in single QDs is expected to be lower than the second-order value \cite{Klimov2000}, it is possible that the appearance of a lower peak in this case is due to a contamination by background fluorescence which contributes more to the background of $g^{(2)}(0)$ than to that of $g^{(3)}(0,0)$\cite{Rundquist2014}. \autoref{fig:g3}(d) summarizes the estimates for the normalized third order correlation, $g^{(3)}(0,0)$, for 19 different QDs. The distribution around zero value demonstrates that third order antibunching is evident in all our measurements.

\section{Quantum image scanning microscopy}
\begin{figure*}[t]
\centering
\includegraphics[width=.7\textwidth]{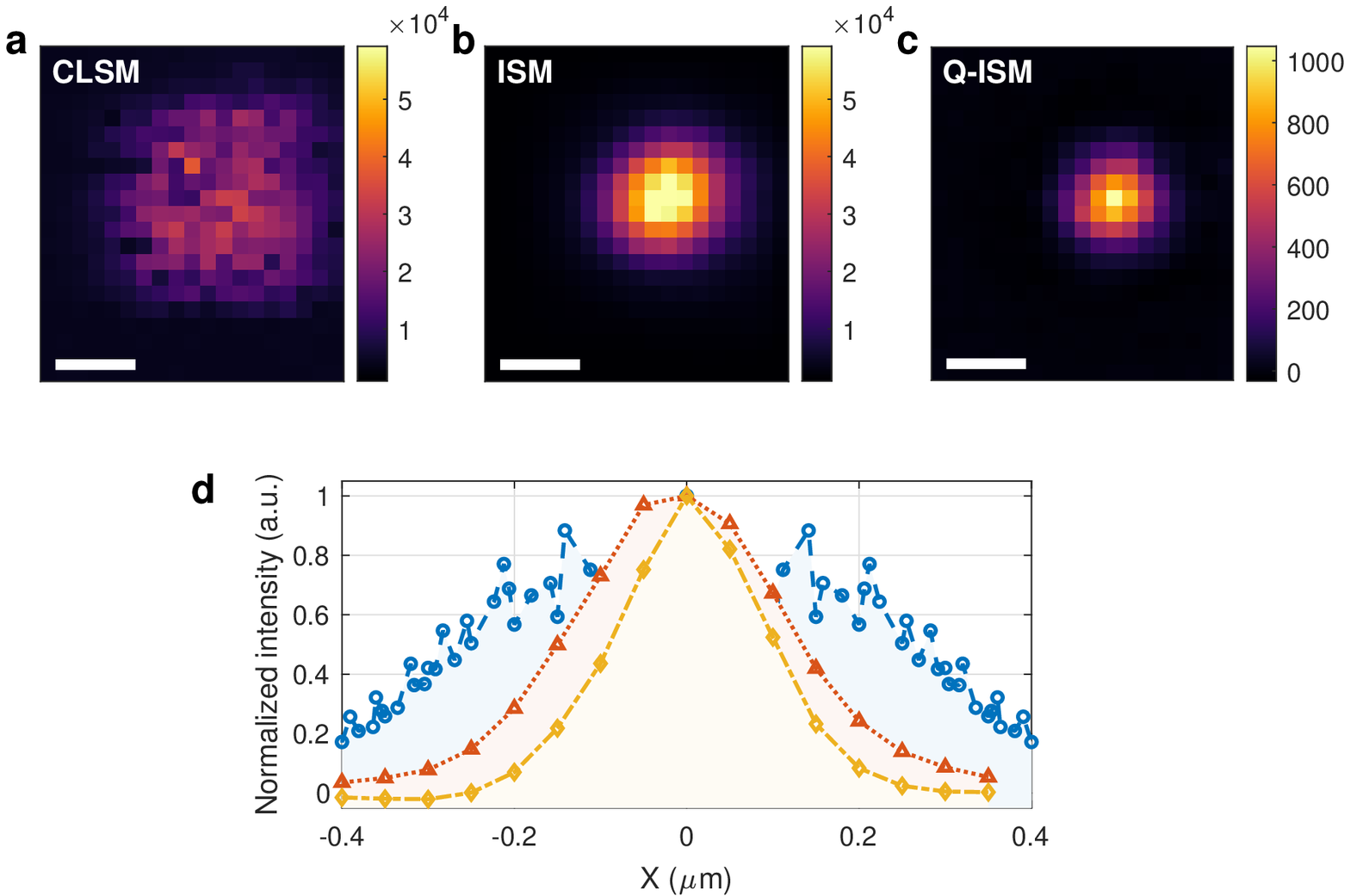}
\caption{\label{fig:qism}\textbf{Narrowing the point spread function with Q-ISM.} A $1\ \si{\micro\meter}\times1\ \si{\micro\meter}$ confocal scan of a single CdSe/CdS/ZnS QD. \textbf{a.}~CLSM image - summing counts over all detectors for each scan position. \textbf{b.}~ISM image - the intensity image generated by each detector is shifted before summation. \textbf{c.}~Q-ISM image - $\Delta G^{(2)}$ for each detector pair is shifted and then summed. \textbf{d.}~Cross-sections for the different analyses: CLSM (blue circles, dashed line), ISM (red triangles, dotted line) and Q-ISM (yellow diamonds, dash-dot line). The values for the CLSM cross section were radially averaged to reduce blinking artifacts. These artifacts do not affect ISM and Q-ISM images. Scale bar: \SI{0.25}{\micro\meter}.}
\end{figure*}

To showcase the system's applicability to quantum optics and imaging science, we demonstrate an implementation of the recently introduced Q-ISM technique\cite{Tenne2018}. This super-resolution scheme utilizes the measurement of quantum correlations in an image scanning microscopy (ISM) architecture. In an ISM scan, the standard pinhole and detector of a confocal microscope are replaced with a detector array\cite{Muller2010}. Merging the scanned images generated by each detector according to their spatial offsets, one can achieve the resolution enhancement of a narrow confocal pinhole while retaining the collection efficiency of a wide confocal pinhole\cite{Sheppard1988,Muller2010}. In Q-ISM, photon detections in each pair of detectors in the array are correlated during the confocal scan, to generate multiple $\Delta G^{(2)}=\langle G^{(2)}_{corr}(\infty)\rangle-G^{(2)}_{corr}(0)$ images. Imaging photon pairs (or rather missing photon pairs at zero time delay) instead of single photons, results in a narrower effective PSF \cite{Tenne2018}. The $\Delta G^{(2)}$ images are merged together as in the ISM technique, to form a super resolved image. This image surpasses the resolution of standard ISM by violating the classical light assumption at the basis of Abbe's diffraction limit\cite{Abbe1873}.

\autoref{fig:qism} shows images constructed from a $\SI{1}{\micro\meter}\times\SI{1}{\micro\meter}$ scan around an isolated QD with a $\SI{50}{\nano\meter}$ step size and a $\SI{200}{\milli\second}$ pixel dwell time. \autoref{fig:qism}(a) shows the result of summing counts of all detectors in the SPAD array for every point in the scan. This is analogous to a confocal laser-scanning microscope (CLSM) with a broad pinhole. Note that due to fluctuations in the fluorescence intensity, some excess noise is obscuring the PSF (see \autoref{app:QD_blink}). \autoref{fig:qism}(b) and \ref{fig:qism}(c) show the ISM and Q-ISM images respectively, attained by the method described above from the same data set. \autoref{fig:qism}(d) presents a comparison of cross sections for the different techniques - showing the PSF narrowing achieved by Q-ISM. The estimated resolution is enhanced by a factor of 1.4 and 1.9 for ISM and Q-ISM images respectively, in agreement with a $\sqrt{2}$ and 2 enhancement factor expected from theoretical considerations\cite{Tenne2018}. To estimate the resolution enhancement we compare the FWHM of the ISM and Q-ISM images shown in \autoref{fig:qism}(b) and \ref{fig:qism}(c) with the FWHM of the CLSM image of a \SI{20}{\nano\meter} diameter fluorescent bead (see \autoref{app:res_estimate}). Further improvement of the resolution can be achieved by deconvolving the Q-ISM image, obtaining a $\sim$2.6 enhanced resolution (see \autoref{app:res_estimate}). The implementation of Q-ISM with an on-chip SPAD array yields a simple, compact and cheap setup as compared to the original fiber bundle camera, highlighting the benefits of using SPAD arrays in quantum imaging schemes.

\section{Discussion}
Compared with previous PNR solutions, the approach of spatial multiplexing with on-chip SPAD arrays has a few advantages and disadvantages. First, the implementation of this approach is relatively simple for the end-user as it does not require cryogenics as is the case for TES and VLPC detectors\cite{Eisaman2011}. In comparison with time multiplexing approaches it is not necessary to manufacture complex and costly fiber systems, typically compatible only with single spatial mode operation\cite{Achilles2003,Fitch2003,Kruse2017}. In fact, the detector array used here can be mounted in the image plane of any confocal microscope making it an attractive solution for quantum based imaging and quantum spectroscopy of nano-structures.

In addition, this method is quite robust to the temporal characteristics of the signal. While time multiplexed detectors require short pulses with a low repetition rate, TES has a temporal resolution of tens of nanoseconds\cite{Lita2008} and both VLPC and TES typically operate up to a \SI{100}{\kilo\hertz} repetition rate\cite{Eisaman2011}; the temporal resolution of CMOS SPAD arrays is limited only by the sub-nanosecond temporal jitter of the SPADs\cite{Villa2012,Burri2016,Burri2017}. The repetition rate, as in any multiplexing method, is limited by the detection saturation (or pile-up) effect\cite{Kruse2017,Miatto2018}; two or more photons can impinge on the same pixel (or time bin) yielding only a single 'click', interpreted as one photon. However, this becomes observable only at \SI{\sim 4}{Mcps} per pixel\cite{Antolovic2018}. Even at the modest number of pixels presented here this allows reliable measurements at up to \SI{\sim 80}{Mcps}, and this can be further enhanced by scaling-up the number of detectors in the array. The number of pixels in this work was chosen to best fit ISM\cite{Castello2015}. However, the compatibility of the manufacturing process of on-chip SPAD arrays with CMOS technology offers an affordable path to scale-up the number of pixels to hundreds, with small changes to the design and performance\cite{Bruschini2019a}. The combination of scalability with low DCR can offer a higher PNR dynamic range than any of the current methods.

Scaling-up the number of pixels beyond a few hundreds on a single-chip is also possible, but dictates a change of the readout scheme. The parallel readout implemented here requires an independent chip pad for each pixel. The typical pad size, 50-\SI{100}{\micro\meter}, results in a trade-off between chip size and pixel number. The two other most frequently used readout schemes are address outputting \cite{Zhang2019} and frame based readout\cite{Ulku2019}, enabling up to $\tfrac{1}{4}$ megapixels at $\sim10^5$ frames per second. Address outputting is usually done at the column level, each column is shared by the pixels that are identified by the row ID. The drawback of this scheme are possible collisions on the column. The frame based readout features memory elements in each pixel. These memories are read out sequentially. Usually, this memory has been a 1-bit memory and hence the time it takes to read the frame becomes the dead time of each pixel. Thus, scaling-up the number of pixels should be accompanied by application-specific requirements, ensuring that the pixel dead time is not significantly increased due to the readout scheme.

A clear disadvantage of the current CMOS SPAD array, with respect to quantum communication protocols, is that the spectral peak of detection efficiency is at \SI{~500}{\nano\meter}\cite{Bronzi2015}; apart of some unique designs\cite{Brida2009}, such devices cannot operate at a telecom wavelength. This disadvantage stems both from the typically shallow junction (p+/n well) depths and the silicon absorption coefficients. Hybrid structures and 3D IC technology are expected to enable detection spectra shifted towards the near infrared. Similarly to other PNR detectors, suffering from inter-detector crosstalk, the subtraction or correction of the crosstalk signal introduces an additional source of noise\cite{Kroger2017}. As discussed in \autoref{sec:ct}, the background term in the zero delay correlation functions is much larger than the signal term. As a result, the main source of error for the estimation of $g^{(2)}(0)$ and $g^{(3)}(0,0)$ is the shot noise on the number of simultaneous photon pairs or triplets respectively (further discussion of the SNR can be found in \autoref{app:SNR}).

In recent years the application of quantum technologies such as quantum sensing, quantum imaging and quantum communication has attracted significant interest. Many of the demonstrated methods in all three areas rely on sensitive multi-port detection of light for the characterization of quantum states of light. A CMOS SPAD array offers a low-noise, compact and cost effective way of performing such measurements. The demonstration of super-resolution imaging based on the concept of Q-ISM shown in this work is an example of one of several concepts that have recently emerged in the field of quantum imaging. These include the enhancement of super-resolution microscopy based on localization microscopy\cite{Israel2017}, structured illumination\cite{Classen2017} and optical centroid measurement\cite{Unternahrer2018,Toninelli2019}, as well as surpassing the classical limits for phase\cite{Ono2013a,Israel2014a} and absorption sensitivity\cite{Brida2010a}. To our knowledge this is the first demonstration of a quantum microscopy modality applying an on-chip detector array. It is a step towards the realization of these methods in a scaled-up, widefield version with an inexpensive detector enabling their application in life-science imaging. Apart from quantum imaging, the few pixel detector used in this work can be used for characterization of nano-scaled sources of quantum light such as quantum dots, organic molecules and solid-state defects. Also, in addition to any quantum state characterization, a SPAD array can in parallel perform the standard measurements of a time correlated single photon counting system, such as lifetime and intensity measurements.

\section{Conclusions}
We have demonstrated the applicability of CMOS SPAD arrays as a scalable, easy to integrate detection array for photon correlation measurements. The implementation of this technique allowed us to measure second and third order photon correlation in the fluorescence of single quantum dots, as well as acquire super-resolved images with the Q-ISM technique. Performing such photon correlation measurements with a simple-to-use and cost-effective detector array can enable widespread use of optical quantum sensing approaches. Scaling up this approach to already existing sub-megapixel CMOS SPAD arrays can pave the way for the application of quantum microscopy in a widefield imaging scenario; thus removing one of the main obstacles in the application of quantum technologies in life-science imaging.

\section*{Funding information}
This work was supported by the Minerva Foundation and the European Research Council (ERC) consolidator grant ColloQuantO.

\section*{Disclosures}
The authors have filed a patent application on the presented method.

\section*{Acknowledgments}
The authors thank Dr. Stella Itzhakov for synthesizing the quantum dots used in this work and Dr. Samuel Burri for the FPGA Spartan 6 readout board.

\bibliographystyle{abbrv}
\bibliography{main}

\clearpage
\begin{center}
    {\Huge Appendices}
\end{center}

\appendix
These appendices describe in further detail the data analysis procedures, detector crosstalk and sample characteristics. Appendices are in order of their reference in the main text: extraction of second and third order photon correlations from the raw data, crosstalk probability temporal stability, quantum dot sample preparation, explicit crosstalk correction terms, quantum dot (QD) fluorescence blinking, resolution enhancement estimate analysis for quantum image scanning microscopy (Q-ISM) and signal to noise ratio.

\section{Raw data analysis} \label{app:analysis}
This section describes the data analysis flow, from digitally logged timestamps to temporal correlations.

The raw data (received from the FPGA) is in the form of a trace of 13 bit integer timestamps with the FPGA's resolution ($t_{clk} = \SI{10}{ns}$). The 13 bit size, results in a `wraparound' of the timestamps every $t_{wrap} = 2^{13} \cdot t_{clk} = \SI{81.92}{\micro s}$. For example, a detection occurring $(2^{13}+1)\cdot t_{clk}$ after the beginning of the measurement will be logged with the timestamp $0$. Hence, the first step is to unravel the wraparounds by adding $t_{wrap}$ between every two non-ascending consecutive timestamps. (This may lead to an artifact whenever count rates for the whole array are similar or below $1/t_{clk} \approx \SI{12}{kcps}$, where wraparounds are 'missed' and detections appear to be closer together. This value is easily surpassed, even when measuring faint sources such as a single QD in it's `grey' state (\autoref{app:QD_blink}). Future FPGA firmware will include more bits per time stamp avoiding this issue all together.)

For the second order correlation ($G^{(2)}$) analysis, pairs of timestamps were binned according to their relative delay. For the $G^{(3)}$ analysis, detections from every photon triplet that arrived within the histogram delay range were randomly assigned the numbers 0, 1 \& 2. The triplets were then binned according to $\tau_1=t_1-t_0$ and $\tau_2=t_2-t_0$. This random assignment is needed, as in contrast to standard `beam-splitting' HBT setups (such as the one depicted in \autoref{fig:intro}(a)), the detectors here are spatially variant. The randomization averages any asymmetry induced by non-uniform illumination of the SPAD array. The results of this step are the as-measured $G^{(2)}_{meas}$ and $G^{(3)}_{meas}$. Following this step, the effects of optical crosstalk are corrected according to the scheme detailed in \autoref{app:ct_cor} to derive $G^{(2)}_{corr}$ and $G^{(3)}_{corr}$.

To assess the degree of antibunching of the measured light we calculate the zero delay time normalized second order correlation, $g^{(2)}(0)$. For this purpose we analyze $G^{(2)}$ at the resolution of the excitation pulse train:
\begin{gather}
    \begin{split}
        \tilde{G}^{(2)}(T) \triangleq \int\limits_{-\frac{T_{pulse}}{2}}^{+\frac{T_{pulse}}{2}} d\tau \cdot G^{(2)}_{corr}(T+\tau) \quad\\ | \quad T=k \cdot T_{pulse}\ ,\ k \in \mathbb{Z}\label{eq:G2_PulseRes}
    \end{split}\\
    \tilde{G}^{(2)}_{DC} = N_p \cdot DCR \cdot T_{pulse} \cdot \frac{N_d-1}{N_d}\\
    g^{(2)}(0) = \frac{\tilde{G}^{(2)}(0)-\tilde{G}^{(2)}_{DC}}{\left\langle\tilde{G}^{(2)}(T)\right\rangle_{T \neq 0}-\tilde{G}^{(2)}_{DC}} \ ,\label{eq:g20Est}
\end{gather}
where $\tilde{G}^{(2)}(T)$ is the second order correlation in pulse period resolution, $T_{pulse}$ is the pulse repetition period (\SI{400}{\nano\second} in the data shown here), $\mathbb{Z}$ is the set of all integers, $\tilde{G}^{(2)}_{DC}$ is the number of detection pairs induced by dark counts within a $T_{pulse}$ delay window; a constant background signal in $\tilde{G}^{(2)}(T)$. $N_p$ is the total number of detected photons, $DCR$ is the dark count rate summed over the entire detector array, $N_d$ is the number of detectors in the array and $\langle\rangle_{T \neq 0}$ is an average over all non-zero $T$'s (often signified as $T=\infty$). In practice the averaging was done over a window of $\abs{T} \leq 400$ (translating to delays up to \SI{\pm160}{\micro\second}). This gave an accurate estimation of $\tilde{G}^{(2)}(\infty)$, due to the averaging, while avoiding correlations at longer time scales (stemming from factors such as light source fluctuations or most of the QD blinking). Indeed, $\tilde{G}^{(2)}(T)$ has an almost constant value in this window (except for, in non-classical sources, $T=-1,0,1$), representing the desired statistics of a memory-less Poissonian source.

Similarly, the normalized third order correlation, $g^{(3)}(0,0)$ is given by
\begin{gather}
    \begin{split}
        \tilde{G}^{(3)}(T_1,T_2) \triangleq \iint\limits_{-\frac{T_{pulse}}{2}}^{+\frac{T_{pulse}}{2}} d\tau_1 \cdot d\tau_2 \cdot G^{(3)}_{corr}(T_1+\tau_1,T_2+\tau_2) \quad\\
        | \quad T_{1,2}=k_{1,2} \cdot T_{pulse}\ ,\ k_{1,2} \in \mathbb{Z}
    \end{split}\\
    \tilde{G}^{(3)}_{DC}(T) =  \tilde{G}^{(2)}(T)\cdot DCR \cdot T_{pulse} \cdot \frac{N_d-2}{N_d}\\
    g^{(3)}(0) = \frac{\tilde{G}^{(3)}(0,0)-\tilde{G}^{(3)}_{DC}(0)}{\left\langle\tilde{G}^{(3)}(T_1,T_2)\right\rangle_{T_{1,2} \neq 0\ \land\ T_1 \neq T_2}-\left\langle\tilde{G}^{(3)}_{DC}(T)\right\rangle_{T \neq 0}}
\end{gather}
where in addition, $\tilde{G}^{(3)}(T_1,T_2)$ is the third order correlation in pulse resolution and $\tilde{G}^{(3)}_{DC}(T)$ is the number of false triplets induced by dark counts within a $T_{pulse}$ delay window (which is now different for $T=0$ and $T \neq 0$).

\section{Crosstalk stability} \label{app:stability}
\begin{figure}
\centering
\includegraphics[width=.9\linewidth]{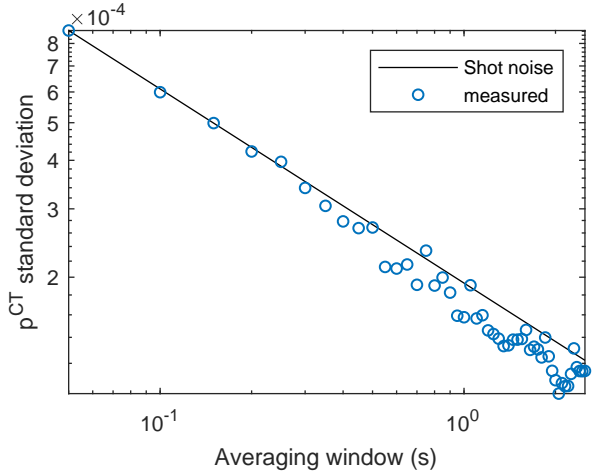}
\caption{\label{fig:ct_stab}\textbf{Crosstalk temporal stability.} Standard deviation of crosstalk probability \textit{versus} the averaging window used to to estimate it (Allan Deviation) over a measurement of \SI{\sim 47}{\second} (blue circles), and the expected Allan Deviation from shot noise (black line). The close agreement indicates stability of the crosstalk probability.}
\end{figure}

The short-term temporal stability of crosstalk probabilities was verified by estimating $p^{CT}_{i,j}$ with various time window durations from a \SI{\sim 47}{\second} measurement of a classical light beam (halogen lamp). \autoref{fig:ct_stab} shows the results of an Allan Deviation analysis of this measurement for two nearest neighbour detectors. A crosstalk probability estimate is calculated for each temporal window period. The probability variance over the entire trace is plotted \textit{versus} the time window duration. The good agreement between the experimental results and a shot noise model for $p^{CT}_{i,j}$ indicates a high degree of short-term temporal stability of the crosstalk process.

While short-term crosstalk probability stability is crucial to the feasibility of the technique, the simplicity of crosstalk calibration as described in the manuscript (a quick calibration that can be done repeatedly), means long-term stability is not a prerequisite. However, the crosstalk probability is stable also over longer time scales, simplifying the measurement procedure. This was assessed by comparing two crosstalk measurements, as described in \autoref{sec:ct}, made at different times. The measured values were very close, with differences distributions matching the expected distribution due to the error in crosstalk probability estimation (shot noise on the number of detected photon pairs), even for measurements taken 8 months apart. Similar difference distributions were observed when comparing two consecutive measurements.

\section{Sample preparation} \label{app:samp_prep}
CdSe/CdS/ZnS QDs were prepared in a colloidal synthesis. Details of the synthesis can be found in the supporting information of Ref \cite{Schwartz2012a}. This resulted in \SI{\sim 6 x 6 x 10}{\nano\meter} nanoparticles with a fluorescence peak around a \SI{617}{\nano\meter} wavelength. For the 'classical light' samples, a drop of a solution of QDs dissolved in Toluene was dropped on a glass coverslip and measured before substantial drying occurred. Samples of isolated QDs used as quantum light sources were prepared by dispersing the same QDs in a 3wt\% solution of poly(methylmetacrylate) (PMMA) in toluene, and spin-coating the solution onto a glass coverslip. 

\section{Crosstalk corrections}
\label{app:ct_cor}
This section describes the explicit crosstalk correction terms for $G^{(2)}$ and $G^{(3)}$.

\subsection{Second order correlation correction}
\label{subsec:2nd_order_corr}
As crosstalk happens only at zero $t_{clk}$ time delay (neglecting the small `leakage' to $\pm t_{clk}$), there is no crosstalk contribution to the second order correlation at non-zero time delays, hence:
\begin{equation}
    G^{(2)}_{corr}(\tau) = G^{(2)}_{meas}(\tau) \quad | \quad \tau \neq 0\ ,
\end{equation}
where $G^{(2)}_{corr}$ is the corrected second order correlation function excluding crosstalk effects and $G^{(2)}_{meas}$ is the as-measured second order correlation function. For zero time delay, the crosstalk contribution from each pixel $i$ would be:
\begin{equation}
    \sum_j n_i\cdot p^{CT}_{i,j} \quad | \quad j \neq i\ ,
\end{equation}
where $n_i$ is the intensity measured in pixel $i$, $p^{CT}_{i,j}$ is the probability of a measured crosstalk event in pixel $j$ given a detection in pixel $i$ and the sum is over all the pixels in the array, except for pixel $i$. The total crosstalk contribution is just the sum over all pixels $i$:
\begin{equation}
    G^{(2)}_{corr}(0) = G^{(2)}_{meas}(0) - \sum_{i \neq j}n_i\cdot p^{CT}_{i,j}\label{eq:G2_CtCorr} .
\end{equation}

\subsection{Construction and correction of Q-ISM images}

Quantum image scanning microscopy (Q-ISM) images are constructed by evaluating the magnitude of the antibunching dip in $G^{(2)}_{corr}$ for every position in the sample. To perform this evaluation, we first correct for the excess amount of simultaneous detection pairs due to optical crosstalk by subtracting a correction term for each scan step and detector pair

\begin{equation}
    G^{(2)}_{i,j\;(corr)}(x,y,0)= G^{(2)}_{i,j\;(meas)}(x,y,0)-(n_i+n_j)\cdot{p^{CT}_{i,j}},
\end{equation}
where $G^{(2)}_{i,j\;(meas)}(x,y,0)$ is the  number of simultaneous detection pairs in detectors $i,j$ during a single scan step at position $[x,y]$. We then temporally integrate $G^{(2)}_{i,j\;(corr)}$ to calculate the second order correlation function at the resolution of the excitation laser pulse train as done in \autoref{eq:G2_PulseRes}.

\begin{multline}
    \tilde{G}^{(2)}_{i,j}(x,y,T) \triangleq \int\limits_{-\frac{T_{pulse}}{2}}^{+\frac{T_{pulse}}{2}} d\tau \cdot G^{(2)}_{i,j\;(corr)}(x,y,T+\tau)\\ \quad | \quad T=k \cdot T_{pulse}\ ,\ k \in \mathbb{Z}\label{eq:G2corr_pulseRes}
\end{multline}

The antibunching scan image for each detector pair can then be calculated using 
\begin{equation}
    \Delta{G^{(2)}_{i,j}}(x,y)= \left\langle\tilde{G}^{(2)}_{i,j}(x,y,T)\right\rangle_{T \neq 0}\label{eq:deltaG2ij} - \tilde{G}^{(2)}_{i,j}(x,y,0)
\end{equation}
where $\langle\tilde{G}^{(2)}_{i,j}(x,y,T)\rangle_{T \neq 0}$ is the average of $\tilde{G}^{(2)}_{i,j}(x,y,T)$ for non-zero pulse delays.

Finally, to construct a Q-ISM image we perform the pixel re-assignment procedure described in reference \cite{Tenne2018}, i.e. shifting the images $\Delta{G^{(2)}_{i,j}}(x,y)$ by a pre-calibrated translation $[\delta x_{i,j}\; ,\delta y_{i,j}]$ and then summing over the detector pair indices $i,j$ ($i\neq{j})$ 

\subsection{Third  order correlation correction}
When correcting the third order correlation function, the quasi-instantaneous nature of the crosstalk feature leads to:
\begin{multline}
    G^{(3)}_{corr}(\tau_1,\tau_2) = G^{(3)}_{meas}(\tau_1,\tau_2)\\ \quad | \quad \tau_1 \neq 0 \land \tau_2 \neq 0 \land \tau_1 \neq \tau_2\ ,
\end{multline}
where $\tau_1$, $\tau_2$ are as defined in \autoref{app:analysis}, $G^{(3)}_{corr}$ is the corrected third order correlation function excluding crosstalk effects and $G^{(3)}_{meas}$ is the as-measured third order correlation function. The lines $\tau_1 = 0$, $\tau_2 = 0$ and $\tau_1 = \tau_2$ (excluding the point $\tau_1 = \tau_2 = 0$) count the number of two coincidental detections, with the third detection at some non-zero time delay from them. For points on these lines the derivation of the second order correlation correction (\autoref{subsec:2nd_order_corr}) holds almost as is:
\begin{multline}
    G^{(3)}_{corr}(\tau_1,\tau_2) = G^{(3)}_{meas}(\tau_1,\tau_2) - \sum_{i\neq j \neq k}n_i\cdot p^{CT}_{i,j}\\ \quad | \quad \tau_1 = 0 \oplus \tau_2 = 0 \oplus \tau_1 = \tau_2\ ,
\end{multline}
where the sum now is over all pixel trios with three unique indices, $k$ being the pixel with a detection at non-zero time delay from the others.

Next, the correction for the point $\tau_1 = \tau_2 = 0$ is estimated from the measured intensities and corrected second order correlations.
\begin{multline}
    G^{(3)}_{corr}(0,0) = G^{(3)}_{meas}(0,0) - \sum_{i\neq j \neq k}\Big[G^{(2)}_{(i,j)\; corr}(0) \cdot p^{CT}_{j,k}\\ + n_i\cdot p^{CT}_{i,j} \cdot \left(p^{CT}_{j,k} + \tfrac{1}{2}p^{CT}_{i,k}\right)\Big] ,
\end{multline}
where $G^{(2)}_{(i,j)\; corr}(0)$ is the corrected second order correlation at zero time delay for detectors $i$ \& $j$ and $n_i$ is the total number of detections in detector $i$. The first, second and third terms in the sum correspond to pathways (i), (ii) and (iii) in Figure4c of the main text respectively. The third term is proceeded by $\tfrac{1}{2}$, as identical elements are present twice in the summation ($(i,j,k)$ \& $(i,k,j)$)

The correction of both the second and the third order correlation are subject to noise as discussed in \autoref{app:SNR}. This may result in some negative values, especially in higher order correlations, where the signal is comparable to the noise induced by crosstalk. These typically average out when integrating over pulse duration as in \autoref{app:analysis}, but might still be present in the final outcome in some fraction of the measurements (leftmost bin in the histogram in \autoref{fig:g3}(d)).

\section{Quantum dot blinking} \label{app:QD_blink}
\begin{figure}[h]
\centering
\includegraphics[width=.9\linewidth]{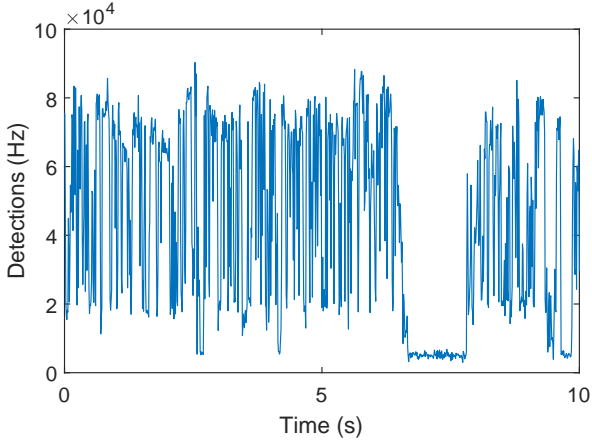}
\caption{\label{fig:blinking}\textbf{Single quantum dot (QD) blinking.} The fluorescence intensity measured from a static single QD with constant illumination presents three distinct fluorescent states: A bright 'on' state, a dark 'off' state and an intermediate intensity 'grey' state. }
\end{figure}

Fluctuations in the luminescence intensity is typical to many types of nano-sized light emitters, including colloidal quantum dots (QDs) such as those used in this work. The digital fluctuation between a dark 'off' state and a bright 'on' state is termed blinking. \autoref{fig:blinking} presents a typical time trace for the fluorescence of a single QD, as measured with the confocal setup described in the main text, employing a CMOS SPAD detector array. Here, we can also observe a short lived, intermediate intensity state commonly referenced in the literature as a 'grey' state. 

These fluctuations manifest as excess noise in the confocal laser scanning microscopy (CLSM) image of a single QD, presented in \autoref{fig:qism}(a) of the main text. The ISM and Q-ISM pixel re-assignment procedures substantially reduce this noise contribution. In these images the value of each pixel is the sum of contributions from different detectors in the array which sample the QD at different times. Therefore, this sum averages the fluorescence intensity of the QD on a time window of a few seconds, smoothing the noise generated by blinking.

At the low detection rates of a dark state, detections may often be separated by more than one wraparound, giving rise to the artifact discussed in \autoref{app:analysis}. Consequently, periods with such low rates will appear shorter than in reality. While this may affect the appearance of the blinking curve it should not have an impact on the antibunching data, especially since both the 'grey' and 'on' state are well above the rate of one detection per wraparound.

\section{Resolution estimate analysis} \label{app:res_estimate}

\begin{figure}
\centering
\includegraphics[width=.9\linewidth]{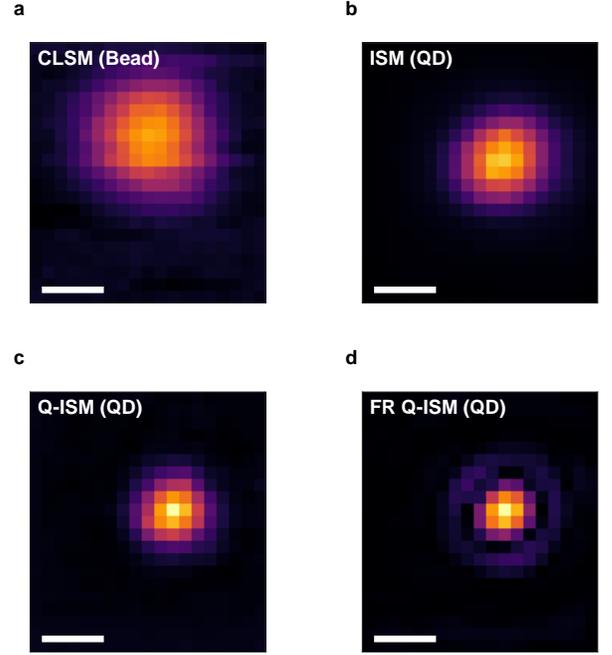}
\caption{\label{fig:resComp}\textbf{Estimating the resolution enhancement in Q-ISM.} Scale bars are \SI{250}{\nano\meter}}
\end{figure}

The following contains a detailed description of the analysis of resolution enhancement estimate provided in the main text. We begin by estimating the resolution obtained with the confocal laser scanning microscopy technique (CLSM) in our experimental setup. For this purpose we use a sparse sample of \SI{20}{\nano\meter} mean diameter fluorescent beads (F8786, Invitrogen), drop casted from solution onto a glass cover slip. By using a non-bliking fluorescent object we avoid additional image noise due to temporal fluctuations in the emitter's QY which may contribute to widening of the analyzed point spread function (PSF). A CLSM image is obtained from the data of a \SI{1x1}{\micro\meter} area scan containing an isolated single bead ($\SI{50}{\nano\meter}$ step size, $\SI{200}{\milli\second}$ pixel dwell time) by summing the number of detections over the entire array for every step in the scan (\autoref{fig:resComp}(a)). This analysis is equivalent to a confocal scan taken with a wide pinhole and its resolution is limited to the width of the laser PSF. The image is fit with a 2D Gaussian function
\begin{equation}
    I(x,y)=A\cdot{e^{-\frac{(x-x_0)^2}{2\cdot\sigma_x^2}}}\cdot{e^{-\frac{(y-y_0)^2}{2\cdot\sigma_y^2}}}+B\ ,
\end{equation}
where $A,B,x_0,y_0,\sigma_x$ and $\sigma_y$ are the fit parameters and $I(x,y)$ is the scan image. The width of the 2D Gaussian can then be defined as $\sigma=\sqrt{\sigma_x^2+\sigma_y^2}$. In the case of the CLSM bead image we obtain $\sigma_{CLSM}\sim\SI{160}{\nano\meter}$. 

While temporal fluctuations in the fluorescence can degrade the CLSM image, both the ISM and Q-ISM analyses overcome short term fluctuations; multiple detectors, sampling the emitters at different times, contribute to the same pixel in the image. As a result, we can construct images of the ISM and Q-ISM PSFs from a scan of a single isolated blinking QD. \autoref{fig:resComp}(b) and \ref{fig:resComp}(c) present the ISM and Q-ISM analysis of such a scan. By repeating the fit procedure described above we obtain  $\sigma_{ISM}\sim\SI{115}{\nano\meter}$ and $\sigma_{\textit{Q-ISM}}\sim\SI{83}{\nano\meter}$ as the width estimate for the ISM and Q-ISM PSFs respectively. In comparison to the CLSM resolution, ISM achieves a $\sim1.4$ resolution enhancement whereas Q-ISM yields a $\sim1.9$ resolution enhancement.

Finally, the resolution of Q-ISM can be further enhanced by performing image deconvolution. By implementing a Wiener filter procedure, as described in the supplementary information of Ref~\cite{Tenne2018}, we construct the Fourier re-weighted (FR) Q-ISM image shown in \autoref{fig:resComp}(d) presenting further narrowing of the PSF. Following the same fit procedure the width of the Gaussian is estimated as $\sigma_{\textit{FR Q-ISM}}\sim\SI{62}{\nano\meter}$, reflecting an enhancement by a factor of $\sim2.6$ compared with the CLSM PSF.

\section{Signal to noise ratio} \label{app:SNR}

Signal-to-noise ratio (SNR) is a key issue for many super-resolution techniques and Q-ISM in particular. Since the antibunching signal is much weaker than the fluorescence signal, even when measured with shot noise limited detectors image acquisition typically requires an order of magnitude longer exposure than a confocal image of the same scene \cite{Tenne2018}.

Observing the calculation of $\Delta\tilde{G}^{(2)}_{i,j}(0)$ in \autoref{eq:deltaG2ij}, we note that since the first term is averaged over many pulse delays it carries a negligible contribution to the noise. As a result, the noise of the antibunching image can be approximated as that of the first term, $G^{(2)}_{i,j}(0)$,

\begin{multline}
        V\left[G^{(2)}_{i,j \; (corr)}(0)\right] = V\left[\tilde{G}^{(2)}_{i,j\;(meas)}(0)\right]\\
        + V\left[(n_i+n_j)\cdot{p^{CT}_{i,j}}\right] \ ,
        \label{eq:varG2_1}
\end{multline}
where $V[..]$ stands for the variance of a random variable. $\tilde{G}^{(2)}_{i,j\;(meas)}$ is the as-measured photon correlation function integrated to the resolution of the laser pulse excitation train, in a similar manner to \autoref{eq:G2corr_pulseRes}. Assuming only a shot noise contribution, both $n_i$ and $\tilde{G}^{(2)}_{i,j\;(meas)}(0)$ variance terms can be estimated as the mean of their corresponding distribution. We note that the mean of latter is a sum of two terms:

\begin{equation}
    \langle\tilde{G}^{(2)}_{i,j\;(meas)}(0)\rangle = 
        \frac{n_i\cdot n_j}{2 N_{pul}}\cdot g^{(2)}(0) + \left(n_i+n_j\right)\cdot{p^{CT}_{i,j}},
        \label{eq:meanG2Meas}
\end{equation}
where $N_{pul}$ is the total number of pulses in an acquisition and $g^{(2)}(0)$ is the actual value of the measured light's second order correlation function (rather than its estimate given by  \autoref{eq:g20Est}). The number of uncorrelated photon pairs was estimated here as $\langle G_{i,j \;(corr)}^{(2)}(\infty) \rangle = \frac{n_i\cdot n_j}{2 N_{pul}}$. 

Plugging the expression from \autoref{eq:meanG2Meas} into \autoref{eq:varG2_1} we obtain 
\begin{multline}
    V\left[G^{(2)}_{i,j \; (corr)}(0)\right] = \frac{n_i\cdot n_j}{2 N_{pul}}\cdot g^{(2)}(0) + \left(n_i+n_j\right)\cdot{p^{CT}_{i,j}}\\
    + \left(n_i+n_j\right)\cdot\left[{p^{CT}_{i,j}}\right]^2
        \label{eq:deltaG2err} \ .
\end{multline}

Since we assumed that ${p^{CT}_{i,j}}$ is a known constant matrix, the third term in \autoref{eq:deltaG2err}, the error on the cross-talk correction term, has a negligible contribution and can be dismissed.

The first term in equation \autoref{eq:deltaG2err} is the standard shot noise for the number of simultaneously measured photon pairs and is therefore only dependent on attributes of the emitter and the detectors' quantum yield (QY). An additional source of noise occurs due to the presence of the cross-talk feature and is highly sensitive to specifics of the detector array. Although on average we factor out the contribution of crosstalk, it still generates additional noise in the measurement of antibunching and the images constructed from the antibunching contrast.

The rest of this section is devoted to calculating the errors in a precise estimation of the antibunching of a single photon emitter as measured with a SPAD array. In this case the correlation function at zero delay is summed over all possible detector pairs
\begin{equation}
    G_{(corr)}^{(2)}(0) = 
        \frac{1}{2} \sum_{i\neq j}{ G^{(2)}_{i,j \; (corr)}(0) }
\end{equation}
and its variance follows
\begin{equation}
    V\left[ G_{(corr)}^{(2)}(0) \right] \approx 
        g^{(2)}(0) \cdot \frac{N_p^2}{2N_{pul}} + 
         6\,\langle p^{CT} \rangle \cdot N_p ,
    \label{eq:varSumG20}
\end{equation}
where $N_p$ is the total number of photons, $\langle p^{CT} \rangle$ is an average value of nearest-neighbor cross-talk probability and the factor of 6 is the number of nearest-neighbors per detector in a hexagonal array.

The performance of a single photon emitter is assessed according to the normalized second-order correlation function at zero delay.
\begin{equation}
    g_{(est)}^{(2)}(0) = \frac{G_{(corr)}^{(2)}(0)}{G^{(2)}(\infty)}
\end{equation}
Since the value at infinite delay can be average over many delays the only error contribution stems from the nominator. Using \autoref{eq:varSumG20} we estimate the error (square root of the variance) as
\begin{equation}
    \delta \left[ g_{(est)}^{(2)}(0) \right] =
        \frac{1}{ \sqrt{G^{(2)}(\infty)} } \,
        \sqrt{ g^{(2)}(0) + 12 \, \frac{\langle p^{CT} \rangle}{p_{ph}} },
        \label{eq:errorEst_g20}
\end{equation}
where $p_{ph} \equiv \frac{N_p}{N_{pul}}$ and $G^{(2)}(\infty)=\frac{1}{2} N_{pul} p_{ph}^2$.

The pre-factor in \autoref{eq:errorEst_g20} is the standard expression of a shot noise limited measurement; the relative error reduces with square root of the number of collected photon pairs. Therefore without cross-talk the error is inversely proportional to the photon detection probability and the square root of the measurement time. The presence of detector cross-talk and introduces additional error. For a perfect single photon emitter ($g^{(2)}(0) = 0$) the error is a product of the standard term and the square root of the ratio between the probability of cross-talk and the probability to detect a photon in a pulse. For the system presented in this work $6\,\langle p^{CT} \rangle \approx 0.01$. 

Considering a typical measurement of a colloidal quantum dot with a \SI{20}{\mega\hertz} laser repetition rate and a signal level of $10^5 \,$ counts per second, a three minute measurement is enough achieve an error below $0.01$ in the estimate of $g^{(2)}(0)$.

\end{document}